\documentclass[aps,twocolumn,prl,groupedaddress,superscriptaddress,amsmath,amssymb,amsthm]{revtex4}
\usepackage{subfigure,hyperref,bbm,times}
\usepackage[T1]{fontenc}
\usepackage{braket}
\usepackage{amsthm}
\usepackage{epsfig}
\usepackage{color}
\usepackage{graphicx}
\usepackage{dcolumn}
\usepackage{bm}

\newcommand{\scal}[2]{\langle#1|#2\rangle}
\providecommand{\openone}{\leavevmode\hbox{\small1\kern-3.8pt\normalsize1}}

\begin{document}

\title{Indistinguishability of elementary systems as resource for quantum information processing}

\author{Rosario Lo Franco}
\email{rosario.lofranco@unipa.it}
\affiliation{Dipartimento di Energia, Ingegneria dell'Informazione e Modelli Matematici, Universit\`{a} di Palermo, Viale delle Scienze, Edificio 9, 90128 Palermo, Italy}
\affiliation{Dipartimento di Fisica e Chimica, Universit\`a di Palermo, via Archirafi 36, 90123 Palermo, Italy}
\author{Giuseppe Compagno}
\affiliation{Dipartimento di Fisica e Chimica, Universit\`a di Palermo, via Archirafi 36, 90123 Palermo,
Italy}

\date{\today }

\begin{abstract}
Typical elements of quantum networks are made by identical systems, which are the basic particles constituting a resource for quantum information processing. Whether the indistinguishability due to particle identity is an exploitable quantum resource remains an open issue. Here we study independently prepared identical particles showing that, when they spatially overlap, an operational entanglement exists which can be made manifest by means of separated localized measurements. We prove this entanglement is physical in that it can be directly exploited to activate quantum information protocols, such as teleportation.
These results establish that particle indistinguishability is a utilizable quantum feature and open the way to new quantum-enhanced applications.  
\end{abstract}

                             

\maketitle

The discovery and exploitation of suitable resources is one of the main aims in quantum information and computation processing \cite{NatNews2017b}. 
Quantum-enhanced technologies often employ identical systems  (e.g., qubits, two-level atoms, photons, electrons, quasiparticles), which are bosons or fermions and constitute the elementary building blocks (particles) of quantum networks \cite{laddReview,exptenphoton,fivequbitexp,RevModPhys.81.1051,benatti2013PRA,QEMreview}. However, while for distinguishable particles an operational framework to exploit their properties is well established by individual operations on each particle \cite{horodecki2009quantum}, for identical particles, which are indistinguishable and individually unaddressable \cite{ghirardi2002JSP,tichy2011essential}, the problem of their direct utilization remains open. This is a main issue for the development of quantum technology relying on identical particles. 

A fundamental feature of composite systems, at the core of quantum algorithms, quantum metrology, quantum key distribution and teleportation \cite{pirandolaReview,vedralReview,eisertPRL2007,QEMreview}, is entanglement. Albeit for systems of distinguishable particles entanglement is well understood \cite{horodecki2009quantum}, for identical particles it has been subject of debate and special treatments to deal with it have been introduced \cite{ghirardi2002JSP,ghirardi2004general,tichy2011essential,benatti2014review,wiseman2003PRL,Li2001PRA,
Paskauskas2001PRA,cirac2001PRA,zanardiPRA,eckert2002AnnPhys,benatti2012PRA,
balachandranPRL,sasaki2011PRA,giulianoEPJD,buscemi2007PRA,tichyFort,LFCSciRep,sciaraSchmidt}. As a matter of fact, for such systems there is no general consensus both on entanglement quantification and if, moreover, useful entanglement may be obtained from the indistinguishability of identical particles. 

Absence of consensus is glaring already in the rather simple situation of independently prepared identical particles, as elucidated in the following. One viewpoint is that, irrespectively of particles overlapping or not, their state is always to be considered entangled, this entanglement being not matter of concern \cite{peresbook,ghirardi2002JSP,ghirardi2004general,tichy2011essential}. Various algebraic operator methods instead assess their state as non-entangled \cite{Li2001PRA,balachandranPRL,benattiOSID2017}. In contrast, when particles spatially overlap, they can be separated by extraction procedures so to get a final state containing useful entanglement \cite{cavalcanti2007PRB,plenio2014PRL}, which is in turn inferred to be that of the initial state \cite{plenio2014PRL}. 
Under the same spatial overlap condition a recent particle-based approach, resorting to the usual notions adopted for nonidentical particles such as von Neumann entropy, assesses the state as entangled \cite{LFCSciRep,sciaraSchmidt}. 

A way out from the impasse, whether or not particle indistinguishability can be a source of useful entanglement, would be provided by an operational framework to harness identical particles. This is the aim of our work. We introduce separated spatially localized (local) measurements to estimate the operational entanglement in systems of independently prepared identical particles under generic spatial overlap configurations. We then verify that this entanglement can be exploited to enable teleportation under local operations and classical communications in a conditional way.

\textbf{Operational entanglement.} 
Let us take two independently prepared nonidentical (distinguishable) qubits, taken as particles, A and B: A is in a state with spatial wave function $\psi$ and internal state (pseudospin) $\uparrow$ and B in a state with spatial wave function $\psi'$ and pseudospin $\downarrow$. The pseudospin states may represent, for instance, components $\pm1/2$ of a spin-$1/2$ particle, two energy levels of an atom, horizontal $H$ and vertical $V$ polarizations of a photon. In the Dirac notation, this two-particle state is $\ket{\Psi}_\mathrm{AB}=\ket{\psi\uparrow}_\mathrm{A} \ket{\psi'\downarrow}_\mathrm{B}\equiv\ket{\psi\uparrow}_\mathrm{A} \otimes \ket{\psi'\downarrow}_\mathrm{B}$. Under individual operations on each particle, that is under local operations and classical communication (LOCC), this state is
manifestly separable and as such unentangled \cite{horodecki2009quantum}. 

We now consider two independently prepared identical particles in a state analogous to the previous one which, in the recent particle-based notation \cite{LFCSciRep}, is
\begin{equation}\label{state1}
\ket{\Psi}=\ket{\psi\uparrow, \psi'\downarrow}.
\end{equation}
This state does not contain particle labels, is completely characterized by the set of one-particle states and cannot be written as a tensor product of the one-particle states, meaning that the vector state must be considered as a whole. Within this formalism, the inner products between states of same dimensionality (two-particle probability amplitude) and of different dimensionality are respectively defined as \cite{LFCSciRep} 
\begin{eqnarray}\label{amp-proj}
\scal{\phi'_1,\phi'_2}{\phi_1,\phi_2}&=&\scal{\phi'_1}{\phi_1}\scal{\phi'_2}{\phi_2}+\eta\scal{\phi'_1}{\phi_2}\scal{\phi'_2}{\phi_1},\nonumber\\
\scal{\phi'}{\phi_1,\phi_2}&=&\scal{\phi'}{\phi_1}\ket{\phi_2}+\eta\scal{\phi'}{\phi_2}\ket{\phi_1},
\end{eqnarray}
where $\phi'_1$, $\phi'_2$, $\phi_1$, $\phi_2$, $\phi'$ are generic one-particle states containing both spatial and pseudospin degrees of freedom and $\eta = \pm 1$ with the upper or lower sign for bosons or fermions. Using these equations, the reduced density matrix of the state of Eq.~(\ref{state1}) can be straightforwardly obtained by partial trace onto an arbitrarily chosen one-particle basis. This permits to determine the von Neumann entropy which quantifies the entanglement of the state in the same way than for nonidentical particles. Now, because identical particles are individually unaddressable, the partial trace must be performed on an arbitrary one-particle basis defined in a given spatial region \cite{LFCSciRep}. 

\begin{figure}[tbp]
\centering
\includegraphics[width=0.46 \textwidth]{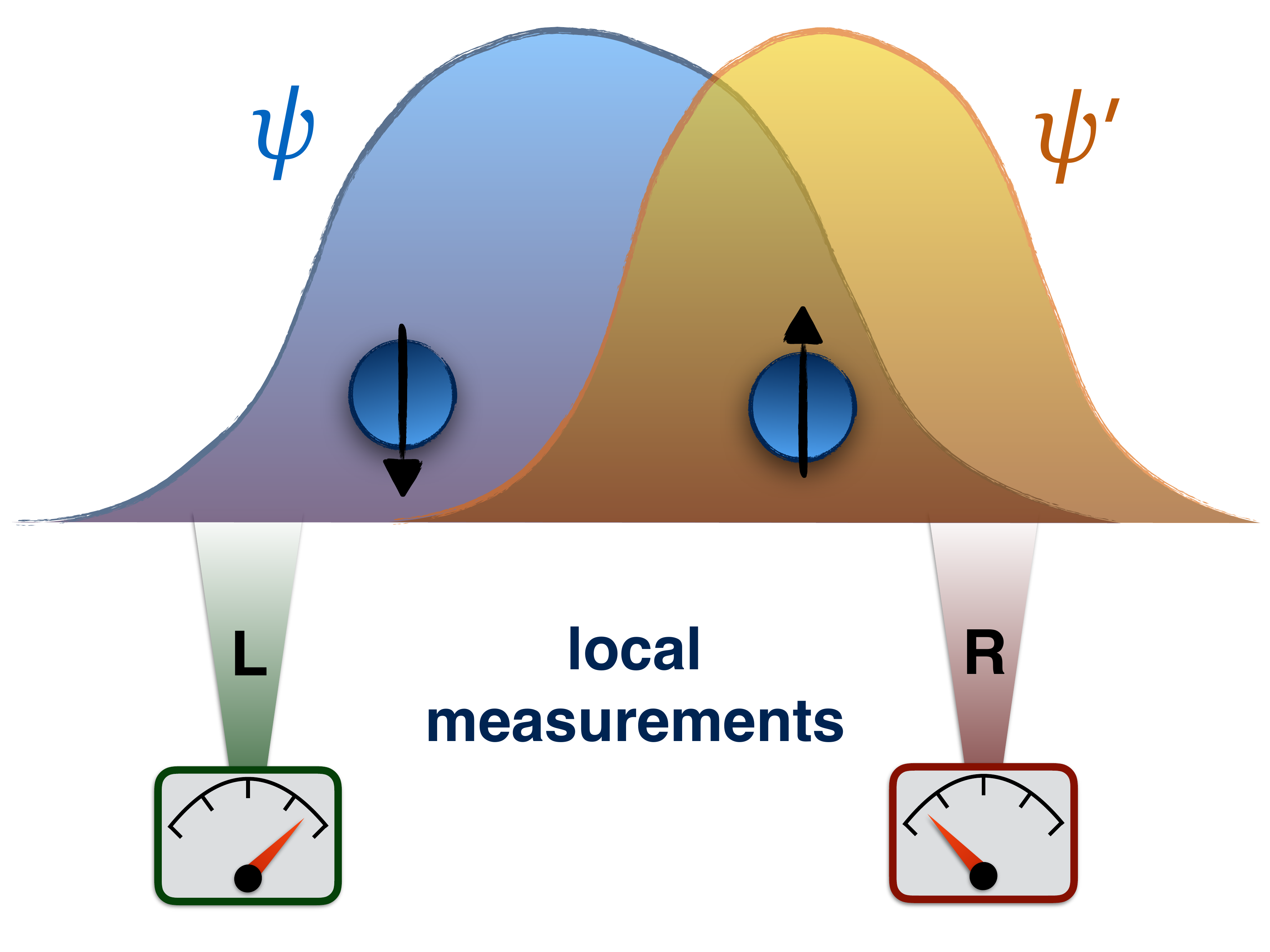}
\caption{{\bf Operational entanglement.} Two identical particles with opposite pseudospins (internal states) have spatial wave functions $\psi$, $\psi'$ with a given degree of overlap. The entanglement between pseudospins is operationally defined by local measurements in two separated localized spatial regions L, R.}
\label{fig:figure1}
\end{figure}

Our first step is to establish an operational framework for determining the entanglement of the state $\ket{\Psi}$ which can be utilized as a tool for quantum information purposes between separated locations. 
We use local measurements of single-particle pseudospin states in separated localized spatial regions L, R, as displayed in Fig. \ref{fig:figure1}. This choice allows one to exclude correlations between the distant regions induced by the measurement process (measurement-induced entanglement \cite{tichy2011essential}). 
So, under this operational procedure, quantum indistinguishability due to particle spatial overlap is necessary for finding nonzero entanglement. We recall that the term ``local'' is here used in the same sense than in quantum field theory, that is a localized region of space, differently from its common meaning for distinguishable particles in quantum information theory, where it indicates an individual particle (particle-locality) \cite{tichy2011essential,benattiOSID2017} irrespective of its spatial distribution. In this sense, the standard operational framework for distinguishable particles based on local operations and classical communication (LOCC) becomes, for indistinguishable particles, based on spatially localized operations and classical communication (sLOCC).

We are ready to investigate the entanglement which comes out within this operational framework. Taking the state $\ket{\Psi}$ of Eq.~(\ref{state1}), we name the probability amplitudes to find a particle in the sites L, R as $l=\scal{\mathrm{L}}{\psi}$, $l'=\scal{\mathrm{L}}{\psi'}$, $r=\scal{\mathrm{R}}{\psi}$, $r'=\scal{\mathrm{R}}{\psi'}$. We perform the (basis-independent) partial trace of $\ket{\Psi}$ on a one-particle basis localized in L and successively project the resulting state onto the site R, finally obtaining the reduced density matrix (see Supplemental Material)
\begin{equation}\label{rho1}
\rho^{(1)}_\mathrm{LR}=\frac{1}{P_\mathrm{L} P'_\mathrm{R} + P'_\mathrm{L} P_\mathrm{R}}(P_\mathrm{L} P'_\mathrm{R} \ket{\downarrow}\bra{\downarrow} + 
P'_\mathrm{L} P_\mathrm{R} \ket{\uparrow}\bra{\uparrow}),
\end{equation}
where $P_\mathrm{L} = |l|^2$, $P'_\mathrm{L} = |l'|^2$, $P_\mathrm{R} = |r|^2$, $P'_\mathrm{R} =|r'|^2$. The von Neumann entropy $S(\rho^{(1)}_\mathrm{LR})=E_\mathrm{LR}(\Psi) =-\mathrm{Tr}(\rho^{(1)}_\mathrm{LR}\log_2 \rho^{(1)}_\mathrm{LR})$ gives the operational entanglement 
\begin{eqnarray}\label{ententropy}
E_\mathrm{LR}(\Psi) &=& 
- \frac{P_\mathrm{L} P'_\mathrm{R} }{P_\mathrm{L} P'_\mathrm{R} + P'_\mathrm{L} P_\mathrm{R}}\log_2 \frac{P_\mathrm{L} P'_\mathrm{R} }{P_\mathrm{L} P'_\mathrm{R} + P'_\mathrm{L} P_\mathrm{R}} \nonumber\\
&-& \frac{P'_\mathrm{L} P_\mathrm{R} }{P_\mathrm{L} P'_\mathrm{R} + P'_\mathrm{L} P_\mathrm{R}}\log_2 \frac{P'_\mathrm{L} P_\mathrm{R} }{P_\mathrm{L} P'_\mathrm{R} + P'_\mathrm{L} P_\mathrm{R}},
\end{eqnarray}
which represents quantum correlations between the pseudospins of the particles observed by local measurements. The amount of this entanglement jointly relies on the probabilities to find the particles in the two localized sites L, R. 

Before analyzing this result in more detail, it is insightful to compare it with the entanglement of the two-particle state, namely $\ket{\Psi_\mathrm{LR}}$, obtained after projecting the pure state $\ket{\Psi}$ of Eq.~(\ref{state1}) onto the two-particle basis $\mathcal{B}_\mathrm{LR}=\{\ket{\mathrm{L}\uparrow,\mathrm{R}\uparrow}, \ket{\mathrm{L}\uparrow,\mathrm{R}\downarrow}, \ket{\mathrm{L}\downarrow,\mathrm{R}\uparrow}, \ket{\mathrm{L}\downarrow,\mathrm{R}\downarrow}\}$.
The projector onto the subspace spanned by the basis $\mathcal{B}_\mathrm{LR}$ is given by the operator  $\hat{\Pi}_\mathrm{LR} = \sum_{\sigma,\tau = \uparrow,\downarrow}\ket{\mathrm{L}\sigma,\mathrm{R}\tau}\bra{\mathrm{L}\sigma,\mathrm{R}\tau}$. The projected normalized two-particle pure state $\ket{\Psi_\mathrm{LR}}$ is then
\begin{eqnarray}\label{PsiLR}
\ket{\Psi_\mathrm{LR}}&=&\hat{\Pi}_\mathrm{LR} \ket{\Psi}/\sqrt{\bra{\Psi}\hat{\Pi}_\mathrm{LR} \ket{\Psi}} \nonumber\\
&=&\frac{lr'\ket{\mathrm{L}\uparrow, \mathrm{R}\downarrow}+
\eta \ l'r\ket{\mathrm{L}\downarrow, \mathrm{R}\uparrow}}{\sqrt{|lr'|^2+|l'r|^2}},
\end{eqnarray}
obtained with probability 
\begin{equation}\label{PLR}
P_\mathrm{LR}=\bra{\Psi}\hat{\Pi}_\mathrm{LR} \ket{\Psi} = P_\mathrm{L} P'_\mathrm{R} + P'_\mathrm{L} P_\mathrm{R} = |lr'|^2+|l'r|^2.
\end{equation}
The entanglement of this state can be then assessed by the standard concurrence for distinguishable particles \cite{horodecki2009quantum} because under local measurements on L and R any of the states of the two-particle basis gives the same probability amplitude of  a separable product state, e.g. $\ket{\mathrm{L}\uparrow,\mathrm{R}\downarrow}\equiv\ket{\mathrm{L}\uparrow}\otimes\ket{\mathrm{R}\downarrow}$ \cite{BLFC2017}. 
Applying the general expression of concurrence for a pure two-particle state (see Supplemental Material) to $\ket{\Psi_\mathrm{LR}}$, one gets $C(\Psi_\mathrm{LR})=2|lr'l'r|/(|lr'|^2+|l'r|^2)$. It is simple to see that the reduced density matrix of the state $\ket{\Psi_\mathrm{LR}}$, after partial trace on either L or R, is equal to $\rho^{(1)}_\mathrm{LR}$ of Eq. (\ref{rho1}). The entanglement of formation \cite{horodecki2009quantum} $E_f = h ((1+\sqrt{1-C^2})/2)$ with $h(x)=-x\log_2 x - (1-x)\log_2 (1-x)$, corresponding to the concurrence $C(\Psi_\mathrm{LR})$ above, coincides with the entanglement entropy of Eq.~(\ref{ententropy}), that is
\begin{equation}
E_\mathrm{LR}(\Psi) = E_f (\Psi_\mathrm{LR}). 
\end{equation}
This equality neatly highlights that, within the established operational framework, the entanglement coming out from the identical particle state $\ket{\Psi}$ of Eq.~(\ref{state1}) is just the standard entanglement of formation of the projected pure state $\ket{\Psi_\mathrm{LR}}$ conditionally obtained as a consequence of operating on the two separated local regions L, R. In this sense, the state $\ket{\Psi_\mathrm{LR}}$ is the distributed resource state between L and R. 

We now discuss these results in terms of spatial overlap of the wave functions. With spatial overlap we mean that the particles can be found in the same region of space and thus that the square moduli of the wave functions ($|\psi|^2$, $|\psi'|^2$) share a region of space where they both differ from zero. 

(i) \textit{No spatial overlap.} Wave functions $\psi$ and $\psi'$ are spatially separated (non-overlapping) and localized, respectively, around L and R, thus $P'_\mathrm{L}=P_\mathrm{R}=0$. Eq.~(\ref{ententropy}) gives no entanglement ($E_\mathrm{LR}(\Psi)=0$); Eq.~(\ref{PsiLR}) gives $\ket{\Psi_\mathrm{LR}}=\ket{\mathrm{L}\uparrow,\mathrm{R}\downarrow}$. As expected, this is due to the fact that the particles are spatially separated and locally addressed.

(ii) \textit{Partial spatial overlap.} Wave functions $\psi$, $\psi'$ partially overlap and different situations arise. When one of the two local measurements on either L or R is performed outside the overlap region (e.g.,  $P'_\mathrm{L}=0$ or $P_\mathrm{R}=0$), Eq.~(\ref{ententropy}) gives zero entanglement ($E_\mathrm{LR}(\Psi)=0$). If both measurements are outside the overlap region, the result of point (i) above is clearly retrieved. Instead, when local measurements occur within the overlap region, entanglement between the particle pseudospins is conditionally obtained, with probability $P_\mathrm{LR}$ given by Eq.~(\ref{PLR}), quantified by Eq.~(\ref{ententropy}) and associated to the state $\ket{\Psi_\mathrm{LR}}$ of Eq.~(\ref{PsiLR}). This analysis evidences that the quantification of the operational entanglement of Eq.~(\ref{ententropy}) is not only an intrinsic property of the state but depends altogether on the structure of the state and on the modality of measurements.  

(iii) \textit{Complete spatial overlap.} The wave functions $\psi$, $\psi'$ exhibit complete spatial overlap when the region where a particle can be found with nonzero probability is the same and, from Eq.~(\ref{ententropy}), the operational entanglement is always nonzero. It is maximum ($E_\mathrm{LR}(\Psi)=1$) when $P_\mathrm{L}=P'_\mathrm{L}=|l|^2$ and $P_\mathrm{R}=P'_\mathrm{R}=|r|^2$ corresponding, from Eq.~(\ref{PsiLR}), to the projected state $\ket{\Psi_\mathrm{LR}^\textrm{max}}=(\ket{\mathrm{L}\uparrow,\mathrm{R}\downarrow}
+ \eta e^{i\gamma}\ket{\mathrm{L}\downarrow,\mathrm{R}\uparrow}) /\sqrt{2}$ with probability, from Eq.~(\ref{PLR}), $P_\mathrm{LR}=2P_\mathrm{L} P_\mathrm{R}=2|lr|^2$. This probability attains its maximum value, $P_\mathrm{LR}^{\textrm{max}}=1/2$, when $|l|^2=|l'|^2=|r|^2=|r'|^2=1/2$. This means that resource states with maximum entanglement can be obtained with a variable efficiency.  

An aspect to be clarified is whether the operational entanglement $E_\mathrm{LR}(\Psi)$, as here defined, is physically sensible. 

\begin{figure}[tbp]
\centering
\includegraphics[width=0.46 \textwidth]{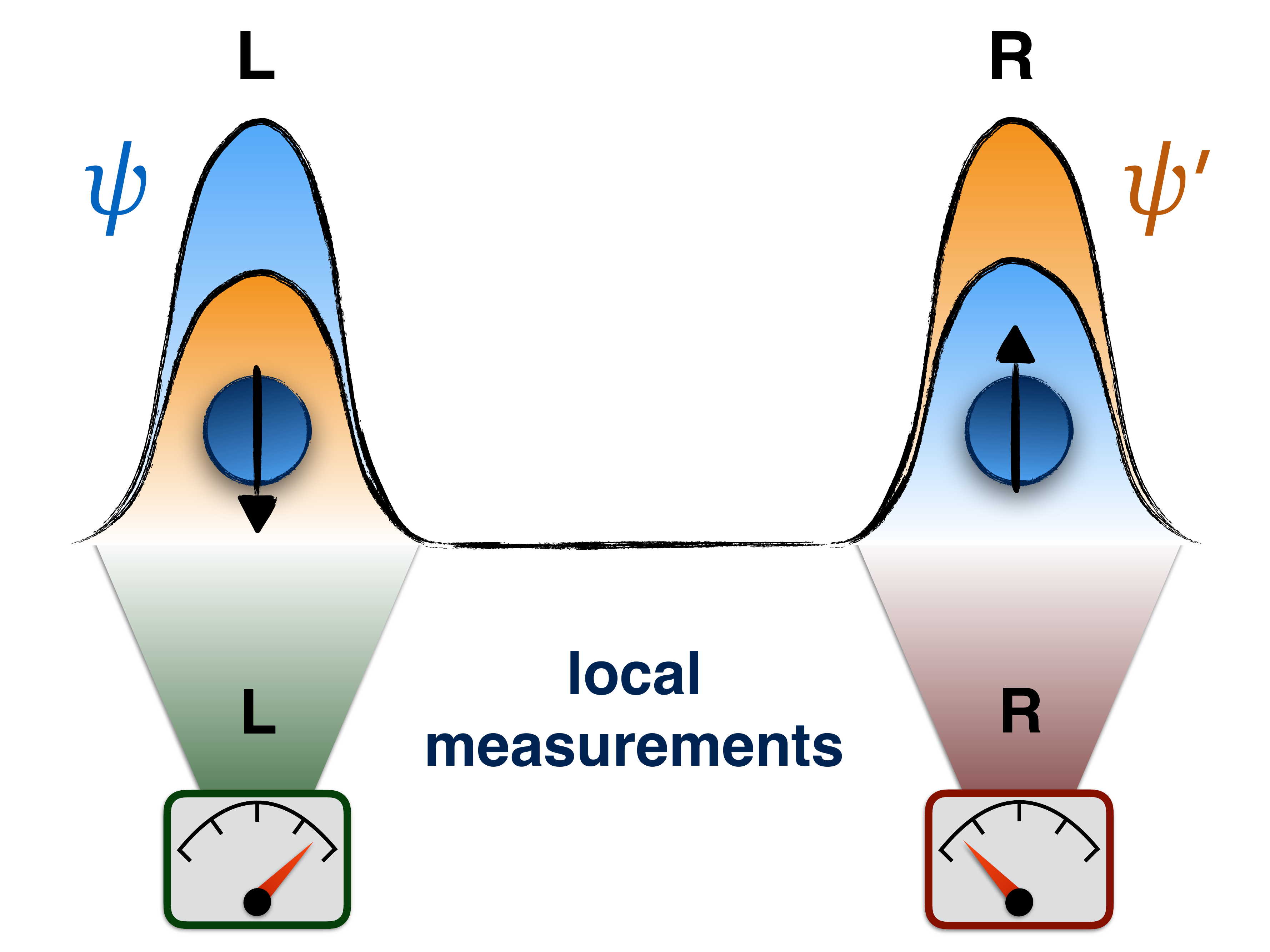}
\caption{{\bf A paradigmatic system.} Two identical particles with opposite pseudospins have overlapping spatial wave functions $\psi$, $\psi'$ peaked in two separated localized regions L, R. Their pseudospins are addressed by local measurements on L and R.}
\label{fig:figure2}
\end{figure}

\begin{figure}[tbp]
\centering
\includegraphics[width=0.48 \textwidth]{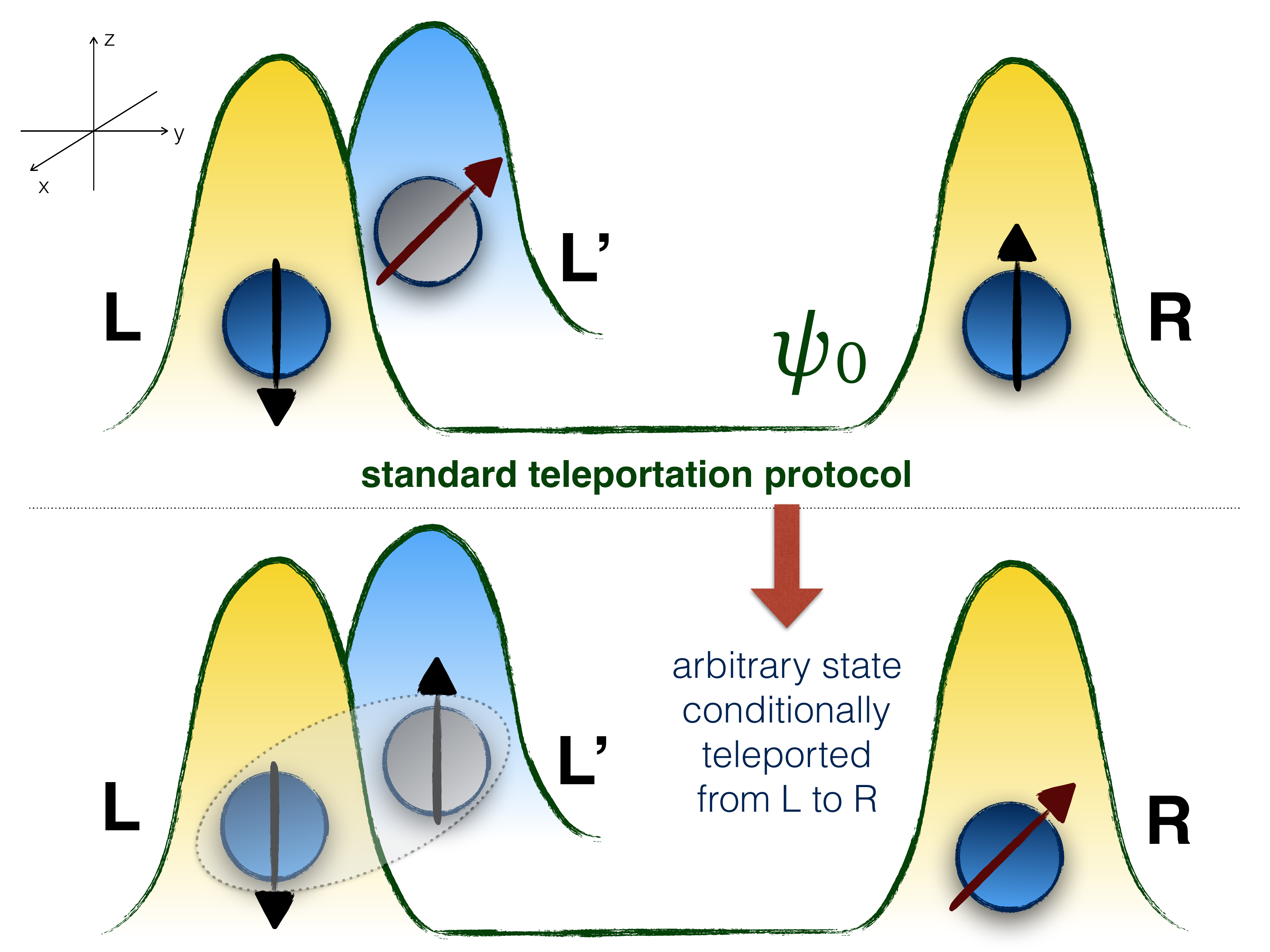}
\caption{{\bf Teleportation scheme.} Two identical particles with opposite pseudospins have the same spatial wave function $\psi_0$ peaked in two separated regions L and R. A third particle, distinguishable from the others, is placed in L$'$ close to L yet separated from it and prepared in an arbitrary pseudospin state. Direct application of the standard teleportation protocol in the two laboratories individuated by (L$'$, L) and R allows the conditional transfer of this arbitrary state from a laboratory to another. The pseudospin of the particle in L$'$ is finally entangled with the pseudospin of one of the two particles in L.}
\label{fig:figure3}
\end{figure}

\textbf{Application.} We shall show this is the case by proving that the entanglement so identified is utilizable in quantum information protocols. 
To this aim, we choose the two particles with spatial wave functions (modes) of Eq.~(\ref{state1}) as
\begin{equation}\label{psispecific}
\ket{\psi}= l \ket{\mathrm{L}} + r \ket{\mathrm{R}},\quad \ket{\psi'}= l' \ket{\mathrm{L}} + r' \ket{\mathrm{R}},
\end{equation} 
where $|l|^2 + |r|^2 = |l'|^2 + |r'|^2 =1$ (see Fig. \ref{fig:figure2}). They are peaked in correspondence to the localized measurement regions (L, R) and are always spatially overlapping, except when either $r=l'=0$ or $l=r'=0$. The two-particle state $\ket{\Psi}$ of Eq.~(\ref{state1}) becomes
\begin{equation}\label{PsiSpecific}
\ket{\Psi}= ll'\ket{\mathrm{L}\uparrow, \mathrm{L}\downarrow} + rr' \ket{\mathrm{R}\uparrow, \mathrm{R}\downarrow}
+ \sqrt{P_\mathrm{LR}}\ket{\Psi_\mathrm{LR}},
\end{equation}
where $\ket{\Psi_\mathrm{LR}}$ and $P_\mathrm{LR}$ are given, respectively, in Eqs.~(\ref{PsiLR}) and (\ref{PLR}). 
From this equation, it is manifest that the distributed resource state $\ket{\Psi_\mathrm{LR}}$ is a part of the global state.    
It is then convenient to take the two wave functions of Eq.~(\ref{psispecific}) such as to maximize both operational entanglement and probability, for instance $\ket{\psi}=\ket{\psi'}=\ket{\psi_0}= (\ket{\mathrm{L}} + \ket{\mathrm{R}})/\sqrt{2}$. With this choice the distributed resource state, available with probability $P_\mathrm{LR}^{\mathrm{max}}=1/2$, is the Bell state $\ket{\Psi_\mathrm{LR}^\textrm{max}}=(\ket{\mathrm{L}\uparrow,\mathrm{R}\downarrow} + \eta \ket{\mathrm{L}\downarrow,\mathrm{R}\uparrow}) /\sqrt{2}$, whose relative phase ($\eta=\pm1)$ is fixed by the particle species (bosons or fermions). A quantum teleportation protocol \cite{BennettTeleportation,pirandolaReview} can be realized as displayed in Fig.~\ref{fig:figure3}. 

Two identical particles are prepared in the state $\ket{\Psi_0}=\ket{\psi_0\uparrow,\psi_0\downarrow}$ and a third particle in an arbitrary state of its pseudospins in the same laboratory where the site L is. This third particle is distinguishable from the others for being either of a different species or identical but separated from the other particles and its state can be written as $\ket{\varphi}_\mathrm{d}= a \ket{\mathrm{L}'\uparrow}_\mathrm{d} + b \ket{\mathrm{L}'\downarrow}_\mathrm{d}$, where L$'$ indicates a site accessible together with L yet separated from the latter (see Fig.~\ref{fig:figure3}).
From Eq.~(\ref{PsiSpecific}) one straightforwardly sees that, excluding the terms when zero or two particles are in mode L, the global initial state conditionally gives with probability $P_\mathrm{LR}^{\mathrm{max}}=1/2$ the state (see Supplemental Material for details)
\begin{eqnarray}\label{teleportation}
\ket{\varphi}_\mathrm{d}\ket{\Psi_0}&\rightarrow
&\frac{1}{2}\left[\ket{\Psi_\mathrm{L'L}^{(\eta)}}\openone_\mathrm{R} + \ket{\Psi_\mathrm{L'L}^{(-\eta)}}\sigma_z^\mathrm{R}
\right. \nonumber \\
&+& \left. \ket{\Phi_\mathrm{L'L}^{(\eta)}}\sigma_x^\mathrm{R} + \ket{\Phi_\mathrm{L'L}^{(-\eta)}}(-i)\sigma_y^\mathrm{R}\right]
\ket{\varphi}_\mathrm{R},
\end{eqnarray}
where $\ket{\Psi_\mathrm{L'L}^{(\eta)}}=(\ket{\mathrm{L'}\uparrow}_\mathrm{d}\ket{\mathrm{L}\downarrow}+
\eta \ket{\mathrm{L'}\downarrow}_\mathrm{d}\ket{\mathrm{L}\uparrow}) /\sqrt{2}$, 
$\ket{\Phi_\mathrm{L'L}^{(\eta)}}=(\ket{\mathrm{L'}\uparrow}_\mathrm{d}\ket{\mathrm{L}\uparrow}
+\eta \ket{\mathrm{'L}\downarrow}_\mathrm{d}\ket{\mathrm{L}\downarrow}) /\sqrt{2}$ are the Bell states between the particle d and one of the two identical particles,  $\openone_\mathrm{R}$ is the identity operator in R, $\sigma_i^\mathrm{R}$ ($i=x, y, z$) are the Pauli matrices and $\ket{\varphi}_\mathrm{R}=a \ket{\mathrm{R}\uparrow} + b \ket{\mathrm{R}\downarrow}$ is the target state teleported in R. Eq.~(\ref{teleportation}) has the structure of the standard teleportation protocol \cite{BennettTeleportation}. The intrinsically conditional recipe to implement the protocol, starting from the initial state, succeeds with probability $P_\mathrm{LR}^{\mathrm{max}}$ and is as follows: (i) the agent Lucy in the laboratory containing L$'$, L performs the Bell measurements and (ii) communicates the outcomes to the agent Rob placed in R who (iii) performs a given operation. If Lucy counts either zero or two particles in L, she tells Rob to reject the procedure; in the other cases, Lucy communicates the outcome $(\ket{\Psi_\mathrm{L'L}^{(\eta)}}$, $\ket{\Phi_\mathrm{L'L}^{(\eta)}}$, $\ket{\Phi_\mathrm{L'L}^{(-\eta)}}$, $\ket{\Psi_\mathrm{L'L}^{(-\eta)}})$ to Rob who makes a corresponding operation $(\openone_\mathrm{R}$, $\sigma_x^\mathrm{R}$, $\sigma_y^\mathrm{R}$, $\sigma_z^\mathrm{R})$ to transform the state of its particle into the desired one. This teleportation protocol, although conditional, is purely quantum since it beats the classical teleportation fidelity threshold $2/3$ \cite{horodecki2009quantum}.
We stress that the probabilistic nature of the process is not related to the use of identical particles but to the locality of measurements and the same would occur for entangled distinguishable particles (see Supplemental Material for details). 
The above results prove the operational entanglement $E_\mathrm{LR}(\Psi)$ associated to independently prepared identical particles is physical. This implies that any other quantum protocol using sLOCC in the same system can be analogously processed. 
The teleportation mechanism here described basically differs from previous ones using identical particles, based on entangled particle number states \cite{marzolino2015,marzolino2016}. 

It is also worth to notice that the state $\ket{\Psi}=\ket{\psi\uparrow, \psi\downarrow}$ as given in Eq.~(\ref{PsiSpecific}) with $\psi=\psi'$ has the very same structure of the final state $\ket{\Psi_\mathrm{ext}}$ which comes out from $\ket{\Psi}$ by the extraction procedure via one-particle tunneling (beam-splitting) transformation  \cite{plenio2014PRL,LFCSciRep}. This fact strenghtens the idea that the extracted entanglement represents the useful entanglement contained in the original state and is not created by the extraction procedure \cite{plenio2014PRL}.

The sLOCC framework within the particle-based approach eventually allows the unambiguous quantification of exploitable identical particle entanglement. In previous works, focus has been on providing schemes, within a linear quantum optics scenario, to generate entanglement exploiting quantum indistinguishability of two identical particles, rather than accessing intrinsic entanglement directly \cite{bose2002}. Tentative steps towards an unambiguous identification of an entangled state of two identical particles have been also reported, formulated as the necessity of having two suitable quantum variables which distinguish the two identical particles \cite{bose2013}. These earlier works may in fact be seen as setting the background for the achievement reached here by sLOCC.

\textbf{Conclusion.} In this work, we have found that indistinguishability of identical elementary systems (particles) can be a resource for quantum information processing. To this aim, we have considered a state of two independently prepared identical particles with generic spatial overlap conditions. We have defined an operational framework using spatially localized operations and classical communication (sLOCC). Under this framework, adopting a particle-based approach to identical particles \cite{LFCSciRep}, we have determined an operational entanglement $E_\mathrm{LR}(\Psi)$ quantified by the von Neumann entropy. It coincides with the standard entanglement of formation as obtained by local operations. The operational entanglement crucially depends on the relative spatial overlap between particle wave functions and among these and measurement regions. 
We have proven that the entanglement measure so obtained can be exploited by sLOCC to conditionally enable quantum teleportation. Particle indistinguishability is therefore revealed as a new source of utilizable entanglement. 
This result, independent of the particle species, occurs when particles are prepared in pairwise orthogonal single-particle states, at variance with the statement that such a state is non-entangled \cite{Paskauskas2001PRA,Li2001PRA,QEMreview}. It motivates studies under the case of non-orthogonal pseudospins and further analyses about the effect of indistinguishability as a source of quantum features other than entanglement, such as coherence \cite{streltsovCoherence}. 

We remark that the particle-based approach here adopted can be seen as complementary to another way to deal with identical particle entanglement, known as algebraic operator approach \cite{benattiOSID2017}. The latter focuses on entanglement between observables while the former defines entanglement in a typical scenario of quantum information theory, so they look in general at different aspects of entanglement. The door seems to be open to show the equivalence of the two approaches when they address the same aspects.

Finally, our study indicates that a basic entangling mechanism can be realized by simply bringing independent identical particles with opposite pseudospins to spatially overlap and then accessing the entanglement by sLOCC measurements. 
In fact, as the LOCC framework permits the unambiguous definition of entanglement of nonidentical particles, the sLOCC framework within the particle-based approach does the same for identical particles. Indeed, this type of operational approach is what is needed to move closer to the spirit of experiments.
This property paves the way to new quantum-enhanced applications in many experimental contexts where identical particles are the elements of quantum networks. For example, Bell experiments \cite{sciarrinoPRA} and teleportation protocol can be realized for photons traveling along overlapping modes with polarizations locally detected at spatially separated places. 
A straightforward implementation can be obtained by a Hanbury Brown and Twiss setup \cite{Brown291} suitably modified with orthogonal polarizers placed before detection. 
Such linear optics realizations may be also reproduced in solid state circuit quantum electrodynamics \cite{PhysRevLett.104.230502,QND2}. Bose-Einstein and fermionic condensates are another natural field of application \cite{QEMreview}, where the particles can be prepared in wells of a lattice and their wave functions adjusted by external parameters like gate voltages, magnetic fields and laser beams \cite{greinerNature,PhysRevLett.117.213001}.

\textbf{Acknowledgments.} 
R.L.F. would like to thank G. De Chiara and F. Sciarrino for discussions. R.L.F. and G.C. also acknowledge A. Castellini, G. Adesso and T. Tufarelli for useful discussions and suggestions.


\appendix

\section{Reduced density matrix}
The desired one-particle reduced density matrix is obtained by partial trace on the initial state $\ket{\Psi}=\ket{\psi\uparrow, \psi'\downarrow}$, following the steps reported in the general approach \cite{LFCSciRep}. The first step consists in performing the partial trace on $\ket{\Psi}$ over an arbitrary one-particle basis localized in L, that is chosen for simplicity to be $\{\ket{\mathrm{L}\uparrow}, \ket{\mathrm{L} \downarrow}\}$, and successively projecting the resulting state onto the site R (the procedure is symmetric with respect to the exchange of L and R). From Eq.~(2) of the main text, one has $\scal{\mathrm{L}\uparrow}{\psi\uparrow,\psi'\downarrow}=\scal{L}{\psi}\ket{\psi'\downarrow}$ and $\scal{\mathrm{L}\downarrow}{\psi\uparrow,\psi'\downarrow}=\eta\scal{L}{\psi'}\ket{\psi\uparrow}$, so that  
\begin{equation}\label{rhoL}
\rho^{(1)}_\mathrm{L}=\mathrm{Tr}_\mathrm{L}\ket{\Psi}\bra{\Psi}= P_\mathrm{L}\ket{\psi'\downarrow}\bra{\psi'\downarrow}
+P'_\mathrm{L}\ket{\psi\uparrow}\bra{\psi\uparrow},
\end{equation}
where $P_\mathrm{L} = |\scal{\mathrm{L}}{\psi}|^2$, $P'_\mathrm{L} = |\scal{\mathrm{L}}{\psi'}|^2$. We are interested in the reduced density matrix corresponding to the separated site R, which is obtained by projecting $\rho^{(1)}_\mathrm{L}$ onto the subspace R by means of the projector $\openone_\mathrm{R}=\ket{R\downarrow}\bra{R\downarrow}+\ket{R\uparrow}\bra{R\uparrow}=\ket{\mathrm{R}}\bra{\mathrm{R}}\otimes (\ket{\downarrow}\bra{\downarrow}+\ket{\uparrow}\bra{\uparrow})$, that represents the identity on the subspace R. 
The (unnormalized) reduced density matrix is then given by 
\begin{equation}
\tilde{\rho}^{(1)}_\mathrm{LR}= \openone_\mathrm{R}\rho^{(1)}_\mathrm{L} \openone_\mathrm{R} = 
P_\mathrm{L} P'_\mathrm{R} \ket{\downarrow}\bra{\downarrow} + P'_\mathrm{L} P_\mathrm{R} \ket{\uparrow}\bra{\uparrow}, 
\end{equation}
where $P_\mathrm{R} = |\scal{\mathrm{R}}{\psi}|^2$, $P'_\mathrm{R} = |\scal{\mathrm{R}}{\psi'}|^2$. The normalized reduced density matrix is finally obtained by dividing for the trace of $\tilde{\rho}^{(1)}_\mathrm{LR}$, as reported in Eq.~(3) of the main text.

\section{Concurrence of a two-qubit pure state}
An arbitrary pure state of two distinguishable particles (qubits) in the standard computational basis $\{\ket{00},\ket{01},\ket{10},\ket{11}\}$ can be written as $\ket{\Phi}=a_{00}\ket{00} +a_{01}\ket{01} +a_{10}\ket{10} + a_{11}\ket{11}$, where the coefficients $a_{ij}$ ($i,j=0,1$) are complex numbers and $\ket{ij}\equiv\ket{i}\otimes\ket{j}$. The general expression of the concurrence quantifying the entanglement of the state $\ket{\Phi}$ is well-known and given by $C(\Phi)= 2 |a_{00}a_{11}-a_{01}a_{10}|$  \cite{horodecki2009quantum}. 

Passing to the L-R basis in our context, we can immediately make the association $\ket{L\downarrow,R\downarrow} \equiv \ket{00}$, $\ket{L\downarrow,R\uparrow} \equiv \ket{01}$, $\ket{L\uparrow,R\downarrow} \equiv \ket{10}$ and $\ket{L\uparrow,R\uparrow} \equiv \ket{11}$. Applying the above general expression of concurrence to the state $\ket{\Psi_\mathrm{LR}}$ of Eq.~(5) in the main text, one obtains $C(\Psi_\mathrm{LR})=2|a_{01}a_{10}|=2|lr'l'r|/(|lr'|^2+|l'r|^2)$.

\section{Teleportation protocol}
The initial identical particle state is $\ket{\Psi_0}=\ket{\psi_0\uparrow, \psi_0\downarrow}$, with $\ket{\psi_0}= (\ket{\mathrm{L}} + \ket{\mathrm{R}})/\sqrt{2}$. The state of the third particle, distinguishable from the others, to be teleported is $\ket{\varphi}_\mathrm{d}= a \ket{\mathrm{L}'\uparrow}_\mathrm{d} + b \ket{\mathrm{L}'\downarrow}_\mathrm{d}$. Omitting the tensor product for simplicity, that is $\ket{\phi}\otimes\ket{\phi'}\equiv \ket{\phi}\ket{\phi'}$, the overall initial state is then
\begin{eqnarray}\label{teleport1}
\ket{\varphi}_\mathrm{d}\ket{\Psi_0}&=&\frac{1}{2} 
\ket{\mathrm{L}'s}_\mathrm{d}\ket{\mathrm{L}\uparrow, \mathrm{L}\downarrow} + \frac{1}{2}\ket{\mathrm{L}'s}_\mathrm{d}\ket{\mathrm{R}\uparrow, \mathrm{R}\downarrow} \nonumber\\
&+& \frac{1}{\sqrt{2}}\ket{\mathrm{L}'s}_\mathrm{d}\left(\frac{\ket{\mathrm{L}\uparrow,\mathrm{R}\downarrow} + \eta \ket{\mathrm{L}\downarrow,\mathrm{R}\uparrow}}{\sqrt{2}}\right),
\end{eqnarray}
where $\ket{s}=a \ket{\uparrow} + b \ket{\downarrow}$ and $\eta = \pm 1$ for bosons and fermions, respectively. This state can be rewritten by inserting, in the first and latter term of Eq.~(\ref{teleport1}), the two-particle identity operator in the basis of the four Bell states $\ket{\Psi_\mathrm{L'L}^{(\pm)}}=(\ket{\mathrm{L'}\uparrow}_\mathrm{d}\ket{\mathrm{L}\downarrow}
\pm \ket{\mathrm{L'}\downarrow}_\mathrm{d}\ket{\mathrm{L}\uparrow}) /\sqrt{2}$ and
$\ket{\Phi_\mathrm{L'L}^{(\pm)}}=(\ket{\mathrm{L'}\uparrow}_\mathrm{d}\ket{\mathrm{L}\uparrow}
\pm \ket{\mathrm{'L}\downarrow}_\mathrm{d}\ket{\mathrm{L}\downarrow}) /\sqrt{2}$. This identity operator is $\openone_{\mathrm{L'}\mathrm{L}}=\ket{\Psi_\mathrm{L'L}^{(+)}}\bra{\Psi_\mathrm{L'L}^{(+)}}+\ket{\Psi_\mathrm{L'L}^{(-)}}\bra{\Psi_\mathrm{L'L}^{(-)}}+\ket{\Phi_\mathrm{L'L}^{(+)}}\bra{\Phi_\mathrm{L'L}^{(+)}}+\ket{\Phi_\mathrm{L'L}^{(-)}}\bra{\Psi_\mathrm{L'L}^{(-)}}$. Notice that the application of $\openone_{\mathrm{L'}\mathrm{L}}$ to the first term of Eq.~(\ref{teleport1}) requires to be divided by 2, that is the number of identical particles present in $\mathrm{L}$, to take into account the correct normalization of the state \cite{LFCSciRep}. We now introduce the notation for products of states of nonidentical and identical particles. We define
$\ket{\phi}_\mathrm{d} \ket{\phi_1,\phi_2} = \ket{\phi}_\mathrm{d} (\ket{\phi_1}\times\ket{\phi_2}) \equiv (\ket{\phi}_\mathrm{d} \ket{\phi_1})\times \ket{\phi_2} +\ket{\phi_1} \times  (\ket{\phi}_\mathrm{d} \ket{\phi_2})$, where we have used the symbol ``$\times$'' to indicate the wedge  product of identical particle states \cite{LFCSciRep}. Utilizing the symmetry of the wedge product for swapping of one-particle states, the second term of previous equation becomes $\ket{\phi_1} \times  (\ket{\phi}_\mathrm{d}  \ket{\phi_2}) = \eta (\ket{\phi}_\mathrm{d} \ket{\phi_2}) \times \ket{\phi_1}= \eta^2 (\ket{\phi}_\mathrm{d} \ket{\phi_1}) \times \ket{\phi_2}$, that then gives
$\ket{\phi}_\mathrm{d} \ket{\phi_1,\phi_2}\equiv 2 (\ket{\phi}_\mathrm{d} \ket{\phi_1})\times \ket{\phi_2}$. Notice that the latter state is in general to be renormalized depending on the initial state. Having introduced this notation, by using the scalar product and projective measurement for nonidentical and identical particles (the latter given in Eq.~(2) of the main text, the overall initial state can be written as
\begin{eqnarray}\label{teleport2}
\ket{\varphi}_\mathrm{d}\ket{\Psi_0} &=& \frac{1}{2}\Big\{ \frac{1}{2\sqrt{2}}\big[\ket{\Psi_\mathrm{L'L}^{(+)}}\times (a\eta \ket{\mathrm{L}\uparrow}+b\ket{\mathrm{L}\downarrow}) \nonumber\\ && + \ket{\Psi_\mathrm{L'L}^{(-)}}\times (a\eta \ket{\mathrm{L}\uparrow}-b\ket{\mathrm{L}\downarrow})\nonumber\\
&&+ \ket{\Phi_\mathrm{L'L}^{(+)}}\times (a \ket{\mathrm{L}\downarrow}+b\eta \ket{\mathrm{L}\uparrow}) \nonumber\\
&&+\ket{\Phi_\mathrm{L'L}^{(-)}}\times (a \ket{\mathrm{L}\downarrow}-b\eta \ket{\mathrm{L}\uparrow}) \big] \Big\}   
\nonumber\\
&& + \frac{1}{2}\ket{\mathrm{L}'s}_\mathrm{d}\ket{\mathrm{R}\uparrow, \mathrm{R}\downarrow} \nonumber\\
&& +\frac{1}{\sqrt{2}}\Big\{ \frac{1}{2}\big[\ket{\Psi_\mathrm{L'L}^{(\eta)}}\times (a \ket{\mathrm{R}\uparrow}+b\ket{\mathrm{R}\downarrow}) \nonumber\\ && + \ket{\Psi_\mathrm{L'L}^{(-\eta)}}\times (a \ket{\mathrm{R}\uparrow}-b\ket{\mathrm{R}\downarrow})\nonumber\\
&&+ \ket{\Phi_\mathrm{L'L}^{(\eta)}}\times (a \ket{\mathrm{R}\downarrow}+b \ket{\mathrm{R}\uparrow}) \nonumber\\
&&+\ket{\Phi_\mathrm{L'L}^{(-\eta)}}\times (a \ket{\mathrm{R}\downarrow}-b \ket{\mathrm{R}\uparrow}) \big] \Big\}.
\end{eqnarray} 
It is now clear that, rejecting the cases when there are zero or two particles in the region L, with probability $1/2$ one can operate with the state within the brace of the last four rows of Eq.~(\ref{teleport2}), reported in Eq.~(10) of the main text with the introduction of the identity operator $\openone_\mathrm{R}$ and the Pauli matrices $\sigma_i^\mathrm{R}$ ($i=x, y, z$) in R. This teleportation protocol, although conditional, is purely quantum since it beats the classical teleportation fidelity threshold $2/3$ \cite{horodecki2009quantum}. In fact, each of the rejected cases, occurring with probability $1/4$, would nevertheless allow teleportation with the classical threshold, so that the total teleportation fidelity of the protocol is: $(1/4) (2/3) + (1/4)(2/3) + (1/2) (1) = 5/6$.

\section{Localized measurements of entangled distinguishable particles}
Let us take the Bell-like state of two distinguishable (nonidentical) particles A, B defined as $\ket{\Psi}_\mathrm{AB}=a\ket{\psi\uparrow}_\mathrm{A} \ket{\psi'\downarrow}_\mathrm{B}+b\ket{\psi\downarrow}_\mathrm{A} \ket{\psi'\uparrow}_\mathrm{B}$ ($|a|^2+|b|^2=1$), whose concurrence is $C(\Psi_\mathrm{AB})=2|ab|$ (see the general expression of concurrence for two-qubit pure states given above). When $\ket{\psi}= l \ket{\mathrm{L}} + r \ket{\mathrm{R}}$ and $\ket{\psi'}= l' \ket{\mathrm{L}} + r' \ket{\mathrm{R}}$, the state 
$\ket{\Psi}_\mathrm{AB}$ becomes
\begin{eqnarray}\label{PsiAB}
\ket{\Psi}_\mathrm{AB} &=& ll' (a\ket{\mathrm{L}\uparrow}_\mathrm{A} \ket{\mathrm{L}\downarrow}_\mathrm{B}+b\ket{\mathrm{L}\downarrow}_\mathrm{A} \ket{\mathrm{L}\uparrow}_\mathrm{B}) \nonumber\\ && 
+ lr' (a\ket{\mathrm{L}\uparrow}_\mathrm{A} \ket{\mathrm{R}\downarrow}_\mathrm{B}+b\ket{\mathrm{L}\downarrow}_\mathrm{A} \ket{\mathrm{R}\uparrow}_\mathrm{B}) \nonumber\\
&& + rl' (a\ket{\mathrm{R}\uparrow}_\mathrm{A} \ket{\mathrm{L}\downarrow}_\mathrm{B}+b\ket{\mathrm{R}\downarrow}_\mathrm{A} \ket{\mathrm{L}\uparrow}_\mathrm{B}) \nonumber\\
&& + rr' (a\ket{\mathrm{R}\uparrow}_\mathrm{A} \ket{\mathrm{R}\downarrow}_\mathrm{B}+b\ket{\mathrm{R}\downarrow}_\mathrm{A} \ket{\mathrm{R}\uparrow}_\mathrm{B}). 
\end{eqnarray}
From this equation it is evident that, due to the individual measurement on distinguishable particles, the degree of entanglement remains fixed ($C(\Psi_\mathrm{AB})=2|ab|$) for any outcome independently of the spatial modes where the measurement is performed. Instead, the probability to obtain an entangled state distributed between two given spatial modes jointly depends on the detection probability amplitudes. This conclusion makes it emerge the conditional appearance of the desired entanglement, for instance, between the particles in the separated regions L, R and the consequent probabilistic nature of a quantum information protocol, like teleportation, involving these regions.

\end{document}